\documentclass[letterpaper, 10 pt, conference]{ieeeconf}  %

\IEEEoverridecommandlockouts                              %
\usepackage{graphicx}
\usepackage{amsmath,amssymb,amsfonts}
\usepackage{mathtools}
\usepackage{float}
\newcommand{\argmin}{\mathop{\mathrm{argmin}}}          %
 
\newtheorem{definition}{Definition}
\usepackage{graphics} %
\usepackage{epsfig} %
\usepackage{color}
\usepackage{caption}
\usepackage{subcaption}
\usepackage{cite}
\usepackage{times} %
\usepackage{url}
\newtheorem{problem}{Problem}

\title{\LARGE \bf
Learning Probabilistic Responsibility Allocations\\for Multi-Agent Interactions
}

\author{Isaac Remy$^{1}$, Caleb Chang$^{1}$, and Karen Leung$^{1,2}$
\thanks{$^{1}$University of Washington, Department of Aeronautics and Astronautics, $^2$NVIDIA.
        {\tt\small \{iremy,cchang26,kymleung\}@uw.edu}.
        I. Remy was supported by the Amazon AI PhD Fellowship. This work was partially supported by the National Science Foundation award under Grant No. 2430686 and 2440861.}%
}

\begin{document}
\newcommand{\responsibility}{\boldsymbol{\gamma}}
\newcommand{\desired}{\mathrm{des}}
\newcommand{\gammavec}{\boldsymbol{\gamma}}
\newcommand{\epsvec}{\boldsymbol{\epsilon}}

\definecolor{DarkGreen}{RGB}{77,167,46}
\definecolor{Red}{RGB}{255,0,0}
\definecolor{Blue}{RGB}{20,85,210}

\maketitle
\thispagestyle{empty}
\pagestyle{empty}

\begin{abstract}
Human behavior in interactive settings is shaped not only by individual objectives but also by shared constraints with others, such as safety. Understanding how people allocate \emph{responsibility}, i.e., how much one deviates from their desired policy to accommodate others, can inform the design of socially compliant and trustworthy autonomous systems. In this work, we introduce a method for learning a probabilistic responsibility allocation model that captures the multimodal uncertainty inherent in multi-agent interactions. Specifically, our approach leverages the latent space of a conditional variational autoencoder, combined with techniques from multi-agent trajectory forecasting, to learn a distribution over responsibility allocations conditioned on scene and agent context. Although ground-truth responsibility labels are unavailable, the model remains tractable by incorporating a differentiable optimization layer that maps responsibility allocations to induced controls, which are available. We evaluate our method on the INTERACTION driving dataset and demonstrate that it not only achieves strong predictive performance but also provides interpretable insights, through the lens of responsibility, into patterns of multi-agent interaction.

\end{abstract}

\section{Introduction}
Designing human-centric autonomous systems requires not only predicting human behavior but also understanding the principles that govern how people interact with one another.
Human interaction is shaped not only by personal objectives but also by shared constraints (e.g., safety) and social norms that govern acceptable behavior. For example, a driver aiming to reach their destination quickly must also follow traffic rules and act in ways that do not endanger others. Understanding how humans balance self-interested goals with the welfare of those around them is key to designing socially-aware autonomous agents.

In this work, we aim to \textbf{codify how humans trade off their own objectives to ensure safe interactions with others}. We build on prior work on \textit{responsibility} in multi-agent interactions \cite{RemyFridovichKeilEtAl2025,CosnerChenEtAl2023}, where responsibility is defined as how much each agent contributes to maintaining shared safety constraints. Unlike previous works, which adopt deterministic models, we propose a \textbf{\textit{probabilistic} responsibility model} derived from real-world human interaction data. Given the inherent multimodal uncertainty of human behavior and the nuanced, context-dependent nature of responsible actions, a probabilistic formulation provides a more faithful and flexible representation of multi-agent responsibility.

\begin{figure}
    \centering
    \includegraphics[width=1.0\linewidth]{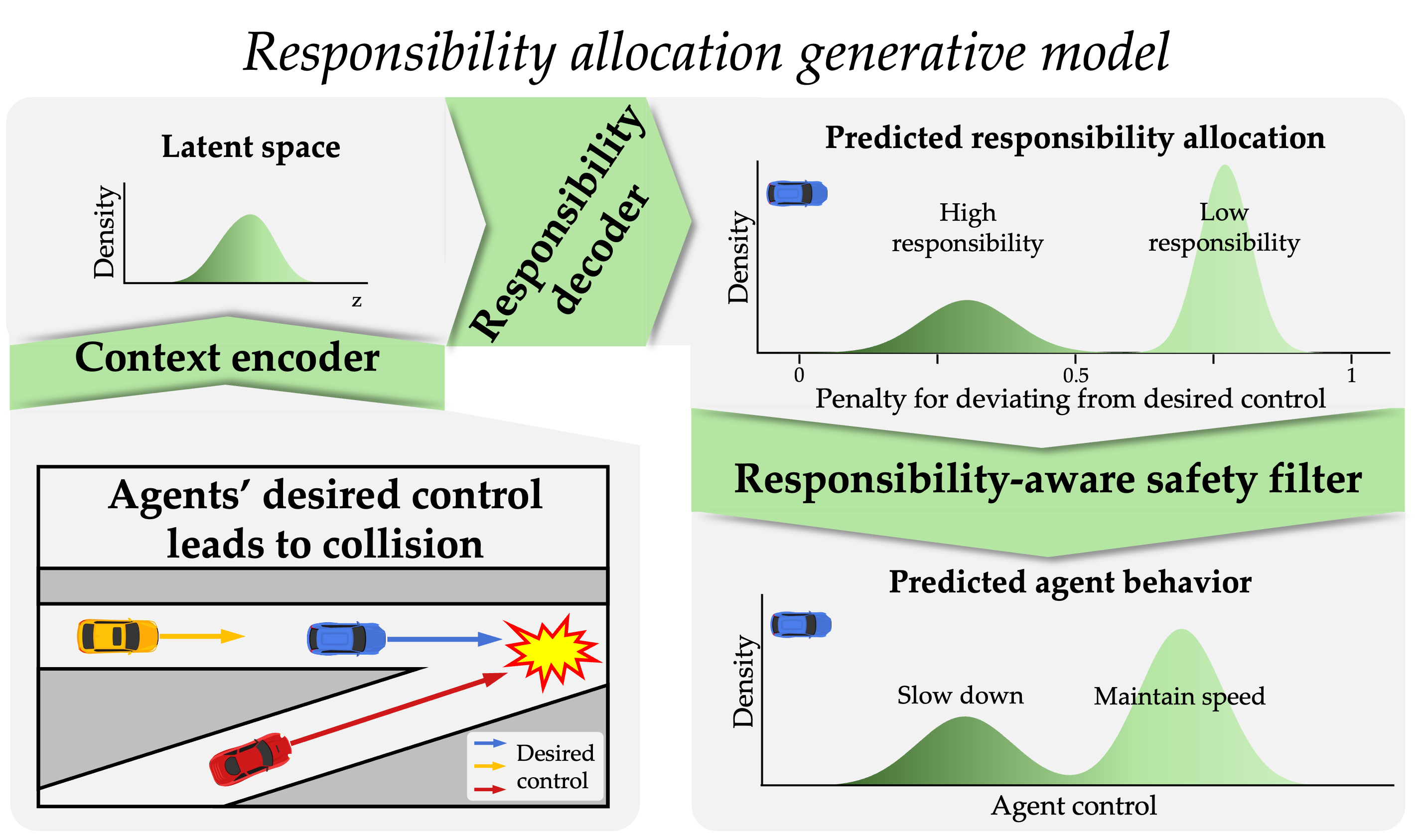}
    \caption{Bottom left: If all agents behave selfishly, executing their desired control, a collision occurs. So who is ``responsible'' for deviating from their desired control? Our responsibility generative model encodes the scene context into a latent space (top left), decodes it to a distribution over responsibility allocations (top right), which pass through a responsibility-aware safety filter, producing a multimodal distribution of agent behaviors (bottom right).}
    \label{fig:hero}
    \vspace{-5mm}
\end{figure}

Specifically, we extend the framework of \cite{RemyFridovichKeilEtAl2025}, which defined an agent’s responsibility as their \textit{willingness to deviate} from their desired behavior to enable safe interaction. That work focused on learning a \textit{deterministic} mapping from the environment to each agent’s responsibility allocation at a single point of time.
In contrast, our probabilistic responsibility allocation model outputs a distribution over responsibility allocations conditioned on current and past environment information.
To achieve this, we leverage techniques in multi-agent trajectory forecasting and generative modeling, which provide tools for handling variable and heterogeneous agents, conditioning on interaction histories, and reasoning over multimodal futures.

\noindent\textbf{Contributions.}  We emphasize that this work is \textit{not} presenting a methodology for trajectory forecasting, but is instead promoting a new way to understand how humans contribute to shared constraints while achieving their individual objectives. Specifically, our contributions are as follows: \textbf{(i)} a formulation of probabilistic responsibility allocations for multi-agent settings that extends prior work while leveraging the strengths of deep generative models and the computational advantages of decentralized control, \textbf{(ii)} a sequence-to-sequence responsibility model borrowing insights from transformer-based multi-agent trajectory forecasting models which can handle numerous complexities, and \textbf{(iii)} experiments demonstrating the efficacy of our probabilistic model on both multimodal synthetic data and real-world driving data. Our code will be released upon paper acceptance.

\section{Related work}
We provide an overview of recent work on socially-aware interaction modeling and trajectory forecasting models.

\subsection{Socially-aware interaction modeling and planning}
Explicitly incorporating the welfare of other agents into an autonomous agent’s planning objective provides an interpretable means of shaping its interactions with others. Prior work utilizing this approach shows the emergence of behaviors described as altruistic \cite{ToghiValienteEtAl2022}, prosocial \cite{GeldenbottLeung2024}, courteous \cite{SunZhanEtAl2018}, cooperative \cite{SchwartingPiersonEtAl2019}, or sympathetic \cite{ToghiValienteEtAl2021}.
These approaches highlight the simplicity of formulating the objective as a linear combination of the agent’s own utility and that of others, where the relative weighting produces markedly different interaction outcomes. For example, \cite{ToghiValienteEtAl2022} shows that altruistic autonomous agents, which assign greater weight to others’ welfare, improve both traffic flow and safety compared to purely egoistic agents.
Rather than specifying the weighting directly, it can also be learned from interaction data \cite{SchwartingPiersonEtAl2019}. In this view, an agent may be assigned a Social Value Orientation (SVO) that characterizes its social preferences (e.g., altruistic, prosocial, or even masochistic).
However, framing the prioritization of others’ welfare as equivalent to ``responsibility'' raises multiple issues:
\textit{(i) Uncertain objectives.} Reasoning at the objective level assumes other agents’ objectives are known. Inferring them through inverse reinforcement learning remains a challenging and open problem, especially in multi-agent settings.
\textit{(ii) Multi-faceted objectives.} An agent’s objective may combine diverse considerations—such as task goals, safety, preferences, and effort—making it unclear what aspect of responsibility is being captured.
\textit{(iii) Decision-making assumptions.} Objective-based reasoning presumes agents are (noisy) optimal decision-makers. While common, this assumption excludes alternative behavior models that directly capture empirical patterns (e.g., behavioral cloning).
Instead of considering how agents value each others welfare, where latent social norms can be hard to disentangle from ``aggressive'' or ``altruistic'' behavior, our responsibility framework prioritizes empirically learning how agents contribute to safety constraints.

\subsection{Trajectory forecasting models}
Recent advances in multi-agent trajectory forecasting \cite{RudenkoPalmieriEtAl2020} have led to a rich set of models and tools for predicting how agents move and interact in complex environments. In particular, there has been significant progress in the autonomous driving domain, fueled by the availability of large open-source datasets \cite{CaesarBankitiEtAl2019,SunKretzschmarEtAl2020}.

\noindent\textbf{Interpretable trajectory forecasting models.} Recent work in trajectory forecasting has increasingly emphasized interpretability, aiming to make predictions more transparent and trustworthy. One direction focuses on disentangling prediction factors such as goals \cite{RhinehartMcAllisterEtAl2019}, intentions \cite{TangMa2025}, and interaction modes \cite{SchmerlingLeungEtAl2018}, allowing users to trace how specific assumptions shape outcomes and inform downstream planning. Another line of work incorporates handcrafted auxiliary features during training to encourage latent variables to align with interpretable concepts. For example, \cite{HsuLeungEtAl2023} introduces ``safety responsibility'' and ``courtesy responsibility'' to ground latent features in a conditional variational autoencoder (CVAE) forecasting model. Collectively, these efforts move beyond simply improving accuracy, aiming instead toward models whose internal logic can be inspected, explained, and aligned with human expectations. Building on this perspective, we design a novel architecture that learns a structured latent embedding representative of responsibility allocations in driving scenarios.

\noindent\textbf{Language-based trajectory forecasting models.} The growing popularity and performance of large language models (LLMs) and vision-language models (VLMs), has led to their increased use in generating, explaining, and reasoning about trajectory predictions \cite{MaoQianEtAl2023,ChenSinavskiEtAl2024,ZhouHuangEtAl2024,XuYangEtAl2025}.
While these results are impressive and promising, language-based explanations are primarily qualitative, computationally expensive, and difficult to leverage directly for \textit{online} decision-making (e.g., trajectory optimization, safety filtering) or \textit{quantitative} evaluation (e.g., crash analysis). 
We view our approach as complimentary to the qualitative nature of LLMs and VLMs; by providing a mathematically grounded and quantitative model of responsibility in multi-agent interactions, we anticipate our method will be easier to integrate into downstream decision-making and evaluation frameworks.
\section{Preliminaries: Defining responsibility }
We provide a brief introduction to Control Barrier Functions (CBFs) and how they are used to define responsibility as introduced in \cite{RemyFridovichKeilEtAl2025}.

\subsection{Control Barrier Functions (CBFs)}
Control barrier functions provide a means for an autonomous system to find safe controls through the principle of forward set invariance. 
Conceptually, CBFs ensure there is a feasible control that prevents a system from approaching a constraint set faster than an allowable rate. A formal definition of CBFs is stated below.

\begin{definition}[Control Barrier Function \cite{AmesGrizzleEtAl2014}]
Consider a dynamical system
$\dot{x} = f(x,u), \quad x \in \mathcal{X} \subseteq \mathbb{R}^n, \; u \in \mathcal{U} \subseteq \mathbb{R}^m$, with locally Lipschitz continuous function $f$.  
A continuously differentiable function $b:\mathcal{X} \to \mathbb{R}$ defines the safe set
$\mathcal{C} = \{x \in \mathbb{R}^n \mid b(x) \geq 0\}$.
The function $b$ is a \emph{Control Barrier Function (CBF)} if there exists an extended class $\mathcal{K}_\infty$ function $\alpha$ such that
\begin{equation}
\sup_{u \in \mathcal{U}} \; \nabla b(x)^T f(x,u) + \alpha(b(x))  \geq 0, \quad \forall x \in \mathcal{C}.
\label{eq:cbf inequality}
\end{equation}
\label{def:CBFs}
\end{definition}
By choosing any controller $k: \mathcal{X} \rightarrow \mathcal{U}$ satisfying the inequality in \eqref{eq:cbf inequality}, then the set $\mathcal{C}$ becomes forward invariant, meaning the system state can stay in $\mathcal{C}$ indefinitely \cite{AmesXuEtAl2017}.

\subsection{Safety filters}
CBFs are commonly used to design a \textit{safety filter}, a module that, if necessary, minimally adjusts a control input to ensure that it satisfies desired safety properties. 
A CBF-based safety filter can be constructed via an optimization problem that minimizes deviations from the desired control input whilst satisfying the CBF constraint \cite{AmesXuEtAl2017}:

\begin{equation}
    \begin{aligned}
        \min_{u\in\mathcal{U}} \:\:& \| u - u_\desired \|_2^2\\
        \mathrm{s.t.} \:\: & \nabla b(x)^T f(x,u) + \alpha(b(x))  \geq 0
    \end{aligned}
    \label{eq:cbf-qp}
\end{equation}
If the dynamics are control-affine, then \eqref{eq:cbf-qp} becomes a quadratic program which can be solved efficiently.

\subsection{Responsibility allocation}
Inspired by the CBF safety filter, which is typically used to filter the control input of a \textit{single} agent, \cite{RemyFridovichKeilEtAl2025} defines responsibility allocations by considering \textit{multi-agent} settings, introducing \textit{weighting parameters} for each agent in the objective.

\begin{definition}[Responsibility allocation]
Consider a \textit{relative} control-affine dynamical system $\dot{\mathbf{x}}_{i,j} = \tilde f(\mathbf{x}_{i,j}) + g(\mathbf{x}_{i,j})\mathbf{u}_{i,j}$ that describes the motion of two agents $i$ and $j$, where $\mathbf{x}_{i,j}$ denotes their relative state (e.g., for a linear single integrator, $\mathbf{x}=[x_j-x_i, y_j-y_i]$) and $\mathbf{u}$ is their concatenated control $\mathbf{u}_{i,j}=[u_i, u_j]$.  Let $b$ be a CBF for the joint system describing the collision set. Then, to ensure safety, the agents $i, j$ have desired control input $\mathbf{u}^\desired_{i,j} = [u_i^\desired, u_j^\desired]$ that is projected into the safe control set via a projection mapping $\mathrm{proj}$. The projection mapping $\mathrm{proj}$ depends on the corresponding joint state $\mathbf{x}_{i,j}$, CBF $b$, class $\mathcal{K}_\infty$ function $\alpha$, and the responsibility allocation vector $\gammavec_{i,j}=[\gamma_i, \gamma_j]$. The responsibility allocation vector determines how much deviation from the desired control each agent is willing to make to satisfy the shared safety constraint described by the CBF inequality. This two-agent projection is used by \cite{RemyFridovichKeilEtAl2025} in their experiments, and we will restate it as Problem \ref{prob:responsible relative CBF safety filter}:
\vspace{1mm}
\begin{problem}[Responsible relative agent CBF safety filter]
{\small
\begin{equation*}
\begin{aligned}
    \mathrm{proj}&(\mathbf{u}_{i,j}^\desired; \mathbf{x}_{i,j}, b, \alpha, \gammavec) \coloneqq  \\
    &\argmin_{\mathbf{u}_{i,j}, \epsilon}  \: \gamma_i \| u_i - u^\desired_i\|_2^2 + \gamma_j \| u_j - u^\desired_j\|_2^2 + \beta_1 \|\mathbf{u}\|_2^2 + \beta_2 \epsilon^2\\
     &\mathrm{s.t.}  \quad \nabla b(\mathbf{x}_{i,j}\left(\tilde f(\mathbf{x}_{i,j}) + g(\mathbf{x}_{i,j})\mathbf{u}_{i,j}\right) + \alpha(b(\mathbf{x}_{i,j})) \geq -\epsilon
     \\
      &\qquad\:\: u_1\in\mathcal{U}_1, u_2\in\mathcal{U}_2\\
      & \qquad\:\: \epsilon \geq 0.
\end{aligned}
\end{equation*}
}
\label{prob:responsible relative CBF safety filter}
\vspace{-2mm}
\end{problem}
\noindent\textit{Remark 1: The desired control can be a simple policy ($u^\desired_i = u^\desired_i(x_i)$) such as maintaining constant velocity, or optimal goal-reaching controls assuming the absence of other agents.}
\noindent\textit{Remark 2: $\beta_1,\beta_2$ are fixed regularization weightings for the control and slack $\epsilon$.}

We want to scale this problem to consider $N$ agents, optimizing over all their relative safety constraints \textit{simultaneously}. Defining a single CBF for multiple agents is nontrivial. Often, the multi-agent CBF constraint is decomposed into pairwise CBFs and the least safe pair is considered, an approach taken in \cite{RemyFridovichKeilEtAl2025} to scale their formulation for $N$ agents; this is flawed as safety is not considered among every pair of interacting agents equally. So, we will write $\mathbf{x} = [x_1, x_2, \dots, x_N]$, $\mathbf{u}=[u_1,u_2, \dots u_N]$, $\gammavec = [\gamma_1, \gamma_2, \dots \gamma_N]$ and define our novel $N$-agent responsibility safety filter as Problem ~\ref{prob:responsible multiagent CBF safety filter}:
\vspace{1mm}
\begin{problem}[Responsible multiagent CBF safety filter]
{\small
\begin{equation*}
\begin{aligned}
    \mathrm{proj}&(\mathbf{u}^\desired; \mathbf{x}, b, \alpha, \gammavec) \coloneqq  \\
    &\argmin_{\mathbf{u}, \epsilon}  \: \sum_{i=1}^N   \left( \gamma_i \| u_i - u^\desired_i\|_2^2 + \beta_1 \|u_i\|_2^2 \right) + \beta_2 \epsilon^2\\
     &\mathrm{s.t.}  \quad \mathbf{G} \mathbf{u} + \mathbf{h} \geq -\epsilon \mathbf{1}
     \\
      &\qquad\:\: u_1\in\mathcal{U}_1,..., u_N\in\mathcal{U}_N\\
      & \qquad\:\: \epsilon \geq 0.
\end{aligned}
\label{eq:responsible multiagent CBF safety filter}
\end{equation*}
}
\label{prob:responsible multiagent CBF safety filter}
\vspace{-2mm}
\end{problem}
\end{definition}
Notice the constraint $\mathbf{G}\mathbf{u} + \mathbf{h}$; this will be our simultaneous CBF safety constraint for $N$ agents; we draw inspiration from \cite{LyuLuoEtAl}, where the CBF is constructed in a ``decentralized'' manner. As opposed to a standard CBF considering a singular ego agent with respect to other agents (treated as obstacles), we construct $C(N,2) := \left(\begin{matrix}
N \\ 2 \end{matrix}\right)$ pairwise CBF constraints so that an agent's control considers each other agent through their relative states and dynamics, as in Problem \ref{prob:responsible relative CBF safety filter}. Specifically, the CBF constraint constants $\mathbf{G} \in \mathbb{R}^{C(N,2)  \times m\cdot N},\, \mathbf{h} \in \mathbb{R}^{C(N,2)}$ for $N$ agents break down as follows:

\begin{equation*}
\begin{aligned}
    \mathbf G = \left[ \begin{smallmatrix} g^{(1)}_{1,\,2} & g^{(2)}_{1,\,2} & 0 & \dots & 0 & 0 \\ g^{(1)}_{1,\,3} & 0 & g^{(3)}_{1,\,3} & \dots & 0 & 0 \\ \vdots \\g^{(1)}_{1, N-1} & 0 & 0 & \dots & g^{(N-1)}_{1,N-1} & 0 \\ g^{(1)}_{1, N} & 0 & 0 & \dots & 0 & g^{(N)}_{1,\,N} \\ 0 & g^{(2)}_{2,\,3} & g^{(3)}_{2,\,3} & \dots & 0 & 0 \\ \vdots \\ 0 & 0 & 0 & \dots & g^{(N-1)}_{N-1,N} & g^{(N)}_{N-1,N} \end{smallmatrix}\right], \,\,\,
    \mathbf{h} = \left[ \begin{smallmatrix}
        h_{1,2} \\ h_{1,3} \\ \vdots \\ h_{1,N-1} \\ h_{1,N} \\ h_{2,3} \\ \vdots \\ h_{N-1,N}
    \end{smallmatrix} \right] %
\end{aligned}
\end{equation*}
Each $g^{(\cdot)}_{i, \,j}$ is a $\mathbb{R}^{1 \times m}$ vector containing the affine terms of the CBF inequality computed for either agent $i$ or agent $j$ using agent $i$ and $j$'s relative dynamics, and each $h_{i,\,j}$ is the corresponding scalar term of the CBF inequality for $i,\,j$. In other words, we are taking $C(N,2)$ CBF constraints from Problem \ref{prob:responsible relative CBF safety filter} and constructing one large CBF inequality for the whole $N$-agent system.

Again, the responsibility allocation vector $\gammavec$ in Problem \ref{prob:responsible multiagent CBF safety filter} penalizes agents' deviation from their desired control differently. The greater $\gamma_i$ is, the more costly it is for agent $i$ to deviate, which is interpreted as them being \textit{less responsible} for contributing to the shared CBF safety constraint.
Thus, given the responsibility allocation vector for agents in a scene, one can identify which agents are more or less responsible for contributing to the shared safety constraint for the group.
The central question becomes: \textbf{\textit{how do we determine the responsibility allocation of agents in a given scene?}}

\section{Problem formulation: Learning responsibility from data}

We aim to learn how responsibility is allocated in multi-agent interactions. Given a dataset of (up to) $N$-agent scenes, let $\mathbf{s}\in\mathcal{S}$ denote the \textit{scene} (including agent states, trajectories, and other relevant information) and $\mathbf{u}=[u_1,\ldots,u_N]$ the observed agent controls. 
Here, we use $\mathbf{s}$ to generalize the concept of ``state'' beyond the concatenation of each agent's dynamical state $\mathbf{x}$. 
Furthermore, we assume we have a dataset, $\mathcal{D} = \{(\mathbf{s}^{(k)}, \mathbf{u}^{(k)})\}_{k=1}^K$.
Our goal is to learn a mapping $\gammavec: \mathcal{S} \rightarrow \mathbb{R}^N$ that predicts a responsibility allocation vector for the $N$ agents in scene $\mathbf{s}$.

\noindent\textbf{Deterministic responsibility mapping.} 
In \cite[Problem 4]{RemyFridovichKeilEtAl2025}, $\gammavec(\cdot)$ was modeled as a deterministic neural network $ \gammavec_\theta $, outputting a fixed allocation for each agent given $\mathbf{s}$. The learning objective was to minimize the discrepancy between the observed controls $\mathbf{u}$ and the controls produced by the responsible multi-agent CBF safety filter (Problem \ref{eq:responsible multiagent CBF safety filter}):
\begin{equation}
    \min_{\theta} \: \mathbb{E}_{(\mathbf{s},\mathbf{u})\sim\mathcal{D}}\left[ \Delta(\mathbf{u}, \mathrm{proj}(\mathbf{u}^{\desired}; \mathbf{s}, b, \alpha, \gammavec_\theta) )  \right],
    \label{eq:deterministic responsibiltiy learning}
\end{equation}
where $\Delta$ is a distance metric (e.g., $\ell_2$). This was tractable since the safety filter optimization is convex and differentiable with respect to\ $\gammavec$. However, a deterministic $\gammavec$ cannot capture multi-modal behaviors that are characteristic to multi-agent interactions.

\noindent\textbf{Probabilistic responsibility mapping.} 
To account for the \textit{multimodal uncertainty} inherent in multi-agent interactions, we instead model $\gammavec$ as a random variable drawn from a conditional distribution $p_\theta(\gammavec \mid \mathbf{s})$. The objective becomes
\begin{equation}
    \min_{\theta} \: \mathbb{E}_{\substack{(\mathbf{s},\mathbf{u})\sim\mathcal{D}, \\ \gammavec \sim p_\theta(\gammavec \mid \mathbf{s})}}\!\left[ \Delta(\mathbf{u}, \mathrm{proj}(\mathbf{u}^{\desired}; \mathbf{s}, b, \alpha, \gammavec)) \right],
    \label{eq:probabilistic responsibiltiy learning}
\end{equation}
where the model learns a distribution over responsibility allocations rather than point estimates. Importantly, this is not a standard generative modeling problem: we do not observe ground-truth responsibility labels, nor are we training a deep model to \textit{directly} learn distribution over controls. Instead, the scene-conditioned distribution over responsibilities $\gammavec$, $p_\theta(\gammavec \mid \mathbf{s})$, is learned indirectly through the induced controls via Problem \ref{prob:responsible multiagent CBF safety filter}. In the next section, we present our approach to learning this distribution using latent generative modeling and differentiable optimization.

\section{Probabilistic Responsibility: A latent variable approach}

The goal is to learn a conditional distribution $p_\theta(\gammavec \mid \mathbf{s})$ so that the induced controls via \eqref{eq:responsible multiagent CBF safety filter} match the observed controls in the dataset.
Given that we do not have labeled data for $\gammavec$ (i.e., unobserved), we use a \textit{generative latent-variable model} to learn $p_\theta(\gammavec \mid \mathbf{s})$ such that its samples satisfy Eq. \ref{eq:responsible multiagent CBF safety filter}.
In particular, we use a Conditional Variational Autoencoder (CVAE) model due to their flexible and representative ability to capture multi-modal distributions \cite{IvanovicLeungEtAl2020}.
First, we give a brief overview of CVAEs, and then outline how we modify them to create a probabilistic responsibility model.

\subsection{Conditional Variational Autoencoders (CVAEs)}
We give a brief introduction to CVAEs; see \cite{Doersch2016} for a more in-depth tutorial.
Given an output variable $y$ (e.g., image) and conditioning variable $x$ (e.g., text), CVAEs seek to model a complex, unknown data distribution $p(y \,|\, x)$.
CVAEs introduce a latent variable $z$, designed to capture hidden factors of variation that are not directly explained by the conditioning variable $x$. The latent space of $z$ intentionally has a lower dimension than the input, encouraging the model to uncover salient features that could aid in interpretability. The standard procedure for learning a CVAE is to optimize the neural network parameters $(\varphi, \psi,\phi)$ that maximize the Evidence Lower Bound (ELBO) loss over the dataset of paired $(x,y)$ data via stochastic gradient descent. The ELBO is defined:
\begin{equation}
\begin{aligned}
\log p(y \mid x) \:\geq \:\: &\mathbb{E}_{q_\varphi(z \mid x, y)} \left[ \log p_\phi(y \mid x, z) \right] - \\
&\mathcal{D}_\mathrm{KL}\left[ q_\varphi(z \mid x, y)\: \middle\|\:\: p_\psi(z \mid x) \right]
\end{aligned}
\label{eq:ELBO}
\end{equation}
Taking gradients of the ELBO is made tractable by using the reparameterization trick \cite{KingmaWelling2022}. $q_\varphi$ represents the \textit{posterior}, which learns $z$ from $x$ and $y$, $p_\psi$ represents the \textit{prior}, which learns $z$ solely from $x$, and $p_\phi$ represents the \textit{decoder}, reproducing $y$ given $x$ and $z \sim q_\varphi(z \mid x, y)$ during training or $z \sim p_\psi(z \mid x)$ during inference. Typically, $p_\psi(z \mid x)$ is chosen to be something easy to sample from (e.g., Gaussian).

\subsection{Responsibility CVAE}
We now describe how we utilize a CVAE model to learn a probabilistic responsibility allocation model. 
The conditioning variable for our problem is $\mathbf{s} = x$ and the output variable is $\mathbf{u}=y$.
Letting $\gammavec = z$ is tempting, but doing this while treating \eqref{eq:responsible multiagent CBF safety filter} as a decoder causes issues to arise.
Representing $p_\psi(z \mid x)$ and $q_\varphi(z \mid x, y)$ as Gaussians makes evaluating the ELBO easier and more efficient (e.g., can use the reparameterization trick, the KL-divergence has a closed-form expression). Thus, if Gaussians are used and we set $\gammavec = z$, this restricts $p(\gammavec \mid \mathbf{s})$ to be a Gaussian, which is not representative of the multimodality inherent in multi-agent interactions. If a non-Gaussian latent distribution is used, computing the ELBO is more challenging, and training becomes less stable and more computationally expensive.

\begin{figure}
    \centering
    \includegraphics[width=\linewidth]{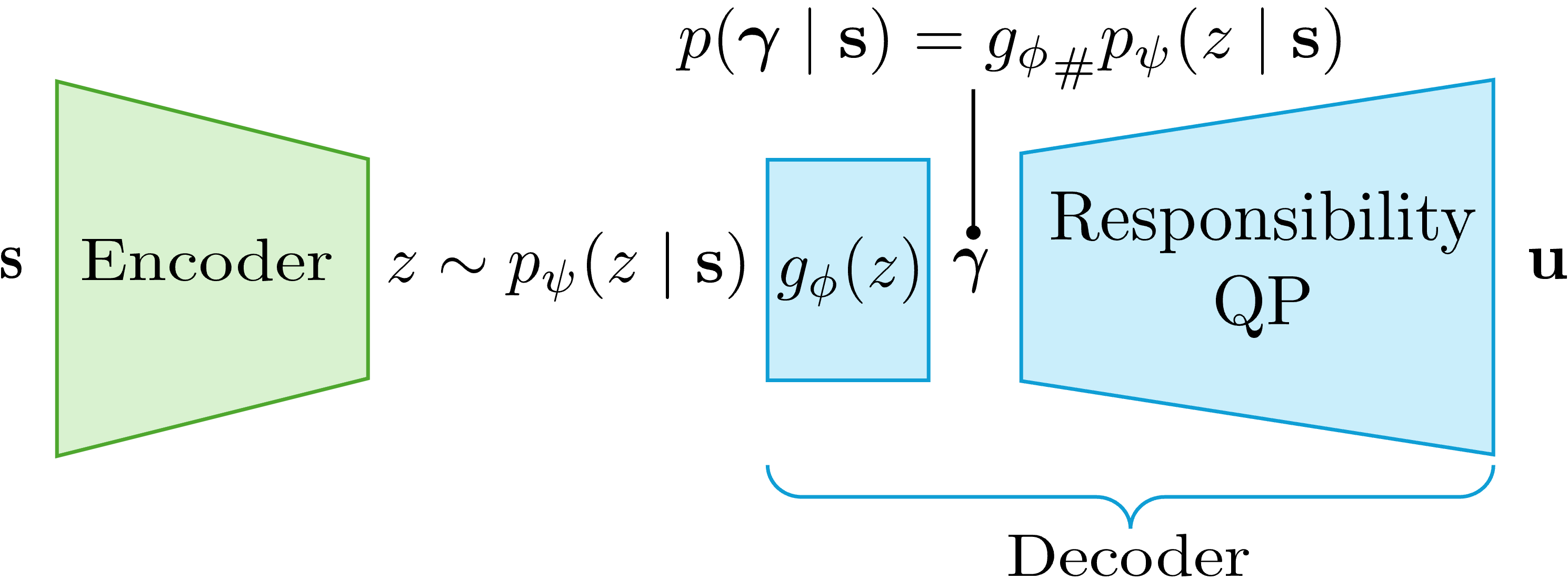}
    \caption{High-level diagram of the responsibility CVAE with continuous latent space. A distribution over responsibility allocation $p(\gammavec \mid \mathbf{s})$ conditioned on scene $\mathbf{s}$ is computed as the pushforward distribution of the latent distribution $p(z \mid \mathbf{s})$ under the initial part of the decoder $g_\phi$. Responsibility allocation values $\gammavec$ are then passed through the responsibility safety filter to produce predicted agent controls $\mathbf{u}$.}
    \label{fig:responsibility cvae diagram}
    \vspace{-5mm}
\end{figure}

We address this issue by keeping the latent variable $z$ as a Gaussian to maintain efficient learning of the CVAE, and treating $\gammavec$ as an \textit{intermediate variable} in the decoder. As illustrated in Fig.~\ref{fig:responsibility cvae diagram}, the decoder consists of several neural network layers, which we denote by $\gammavec = g_\phi(z)$, and a differentiable optimization layer defined by $\mathrm{proj}(\cdot)$, and the final output will be $\hat{\mathbf{u}}$, our predicted control of all agents conditioned on the scene. 
Assuming a Gaussian model around our predicted control given some $z$ and $\mathbf{s}$, we have,
\begin{equation}
    p(\mathbf{u} \mid \mathbf{s}, z) = \mathcal{N}(\mathbf{u} \mid \mathrm{proj}(\mathbf{u}^\desired; \mathbf{s}, b, \alpha, g_\phi(z)), \sigma^2I).
\end{equation}
\noindent\textit{Note}: The variance $\sigma$ can be learned as a function of $z$, or could be constant if desired.

\noindent The training objective for the responsibility CVAE becomes becomes Problem \ref{prob:final-elbo-loss}:
\begin{problem} [Responsibility CVAE objective]
\begin{equation*}
    \begin{aligned}
    \min_{\phi, \psi, \varphi} \: &\mathbb{E}_{(\mathbf{s},\mathbf{u})\sim\mathcal{D}} \Bigl[  \mathcal{D}_\mathrm{KL}\left( q_\varphi(z \mid \mathbf{s}, \mathbf{u})\: \middle\|\:\: p_\psi(z\mid \mathbf{s}) \right)  - \\
    & \mathbb{E}_{q_\varphi(z \mid \mathbf{s}, \mathbf{u})} \left[ \log \mathcal{N}(\mathbf{u} \mid \mathrm{proj}(\mathbf{u}^\desired; \mathbf{s}, b, \alpha, g_\phi(z)), \sigma^2I) \right] \Bigr]
\end{aligned}
\label{eq:final-elbo}
\end{equation*}
\label{prob:final-elbo-loss}
\end{problem}

\noindent\textbf{Obtaining $p(\gammavec \mid \mathbf{s})$ with \textit{continuous} latent space.} 
Given that $p(z \mid \mathbf{s})$ is a Gaussian and therefore easy to sample from, and $g_\phi(\cdot)$ is a set of neural network layers, then $p(\gammavec \mid \mathbf{s})$ is the \textit{pushforward distribution} of $p(z \mid \mathbf{s})$ under $g_\phi$.
To sample from $p(\gammavec \mid \mathbf{s})$, we simply sample from $p(z \mid \mathbf{s})$ and pass the samples through $g_\phi$.

\noindent\textbf{Obtaining $p(\gammavec \mid \mathbf{s})$ with \textit{discrete} latent space.}
In the case of a discrete latent \cite{JangGuEtAl2017}, we can model $p(\gammavec \mid \mathbf{s})$ as a Gaussian mixture model (GMM) where each mode corresponds to each discrete latent vector, and the mean and variance of that mode are computed from passing $z$ through neural network layers.

\section{Sequence-Conditioned Responsibility Prediction}
\label{sec:transformer model}
Given a sequence of past agent states, rather than the CVAE purely predicting possible possible future agent trajectories like in \cite{YuanWengEtAl2021}, we want it to \textit{predict responsibilities that could induce those trajectories}. Importantly, we want the encoder and decoder in our model to handle large, complex scenes where there are a large number of agent trajectories where the number of agents may vary with time. In designing a CVAE capable of working with this type of data, we are inspired by AgentFormer \cite{YuanWengEtAl2021}, which serves as a guide for constructing a CVAE that can handle large quantities of time-sequenced agent states. The transformer architecture is very amenable to our problem, so we take our learning objective from \eqref{eq:final-elbo}, and use it to train our transformer-based responsibility CVAE. We specifically use AgentFormer due its capacity to handle agents moving in and out of the scene, its masking mechanism that incentivizes the learning of relationships between agents, and its permutation invariance (i.e., agent ordering does not matter). In this section, we discuss our generative responsibility model in the context of sequence prediction.%

\subsection{Multi-Agent Trajectory Representation}
First we need to define the inputs and outputs of our responsibility transformer. We assume access to agent state trajectories of up to history length $T \in \mathbb{N}$, with $1, 2, \dots, N$ agents of possibly heterogeneous types (pedestrians, cars, bicycles, etc.). At a time $t$ (with $1 \leq t \leq T$), each agent $i$'s state $s^{(t)}_i \in \mathbb{R}^n$ contains an integer denoting its type, as well as position, velocity, and heading values. To make these multi-agent trajectories usable for a sequence-to-sequence transformer model, we concatenate all the agent states at each time into a two-dimensional sequence, $\mathbf{s}_{\text{past}}$, where:
\begin{align*}
    \mathbf{s}_{\text{past}} = [s^{(1)}_1, \,\dots, \, s^{(1)}_N, \, s^{(2)}_1, \dots, \, s^{(2)}_N, \, \dots, s^{(T)}_1, \dots, \, s^{(T)}_N]^{\top}
\end{align*}
For example, for a fixed $N$ agents in a scene, $\mathbf{s}_{\text{past}} \in \mathbb{R}^{(N \cdot T) \times n}$. However, in reality, different agents are moving in and out of the scene at any given time, so the length of any given $\mathbf{s}_{\text{past}}$ may vary for a fixed history length $T$.

\begin{figure*}[t]
\begin{subfigure}{0.33\textwidth}
  \centering
  \includegraphics[width=\linewidth]{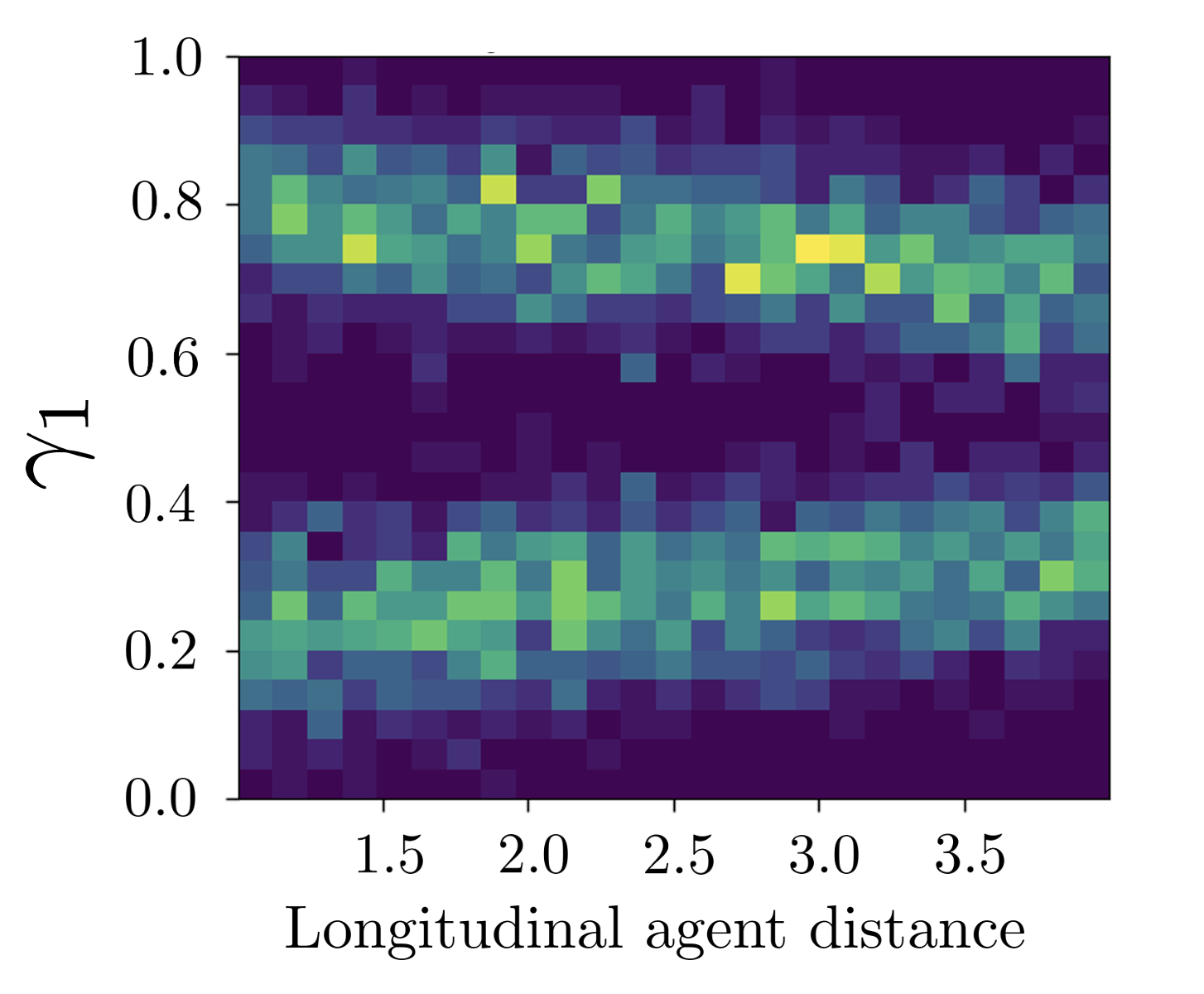}
  \vspace{-7mm}
  \caption{Ground truth $p(\gamma_1 \mid \mathbf{s})$}
  \label{fig:synthetic true gamma x}
\end{subfigure}%
\begin{subfigure}{0.33\textwidth}
  \centering
  \includegraphics[width=\linewidth]{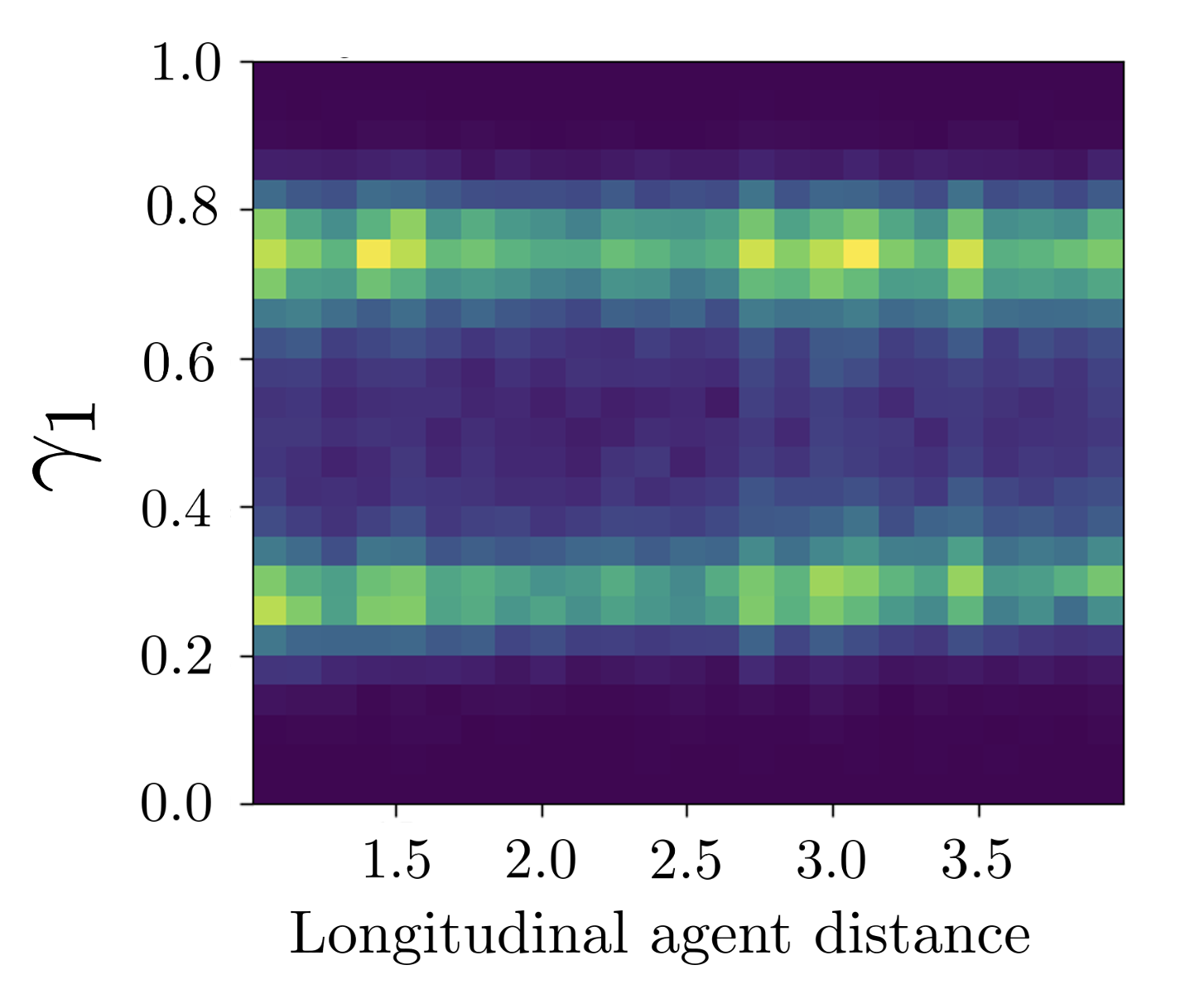}
  \vspace{-7mm}
  \caption{Estimated $p(\gamma_1 \mid \mathbf{s})$ with Gaussian $\mathbf{z}$}
  \label{fig:synthetic continuous gamma x}
\end{subfigure}
\begin{subfigure}{.33\textwidth}
  \centering
  \includegraphics[width=\linewidth]{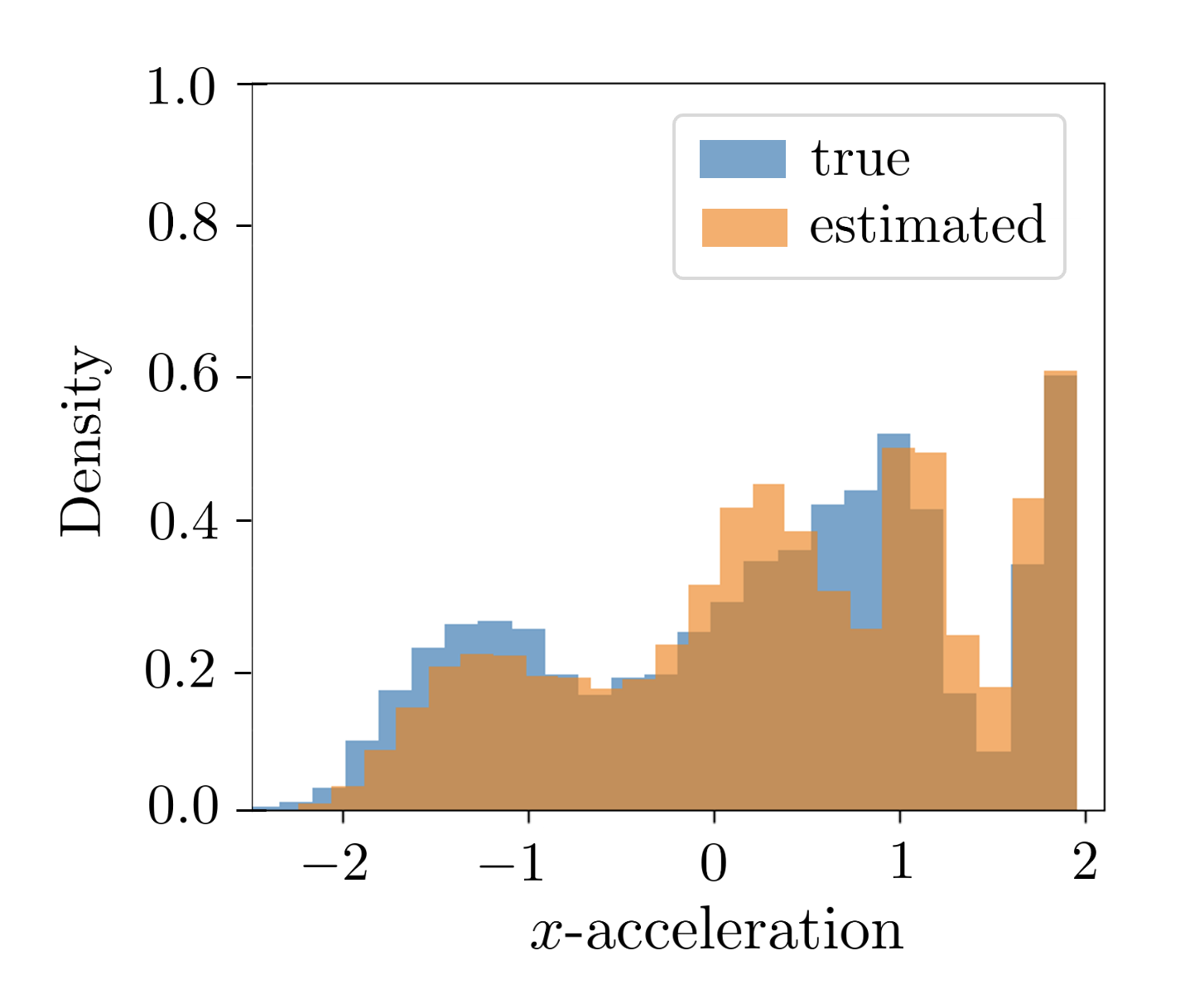}
  \vspace{-7mm}
  \caption{True and estimated $u_{x,1}$ with Gaussian $\mathbf{z}$}
  \label{fig:synthetic u}
\end{subfigure}
\caption{Learning synthetic $p(\gammavec \mid \mathbf{s})$. \textbf{(a)} The ground truth distribution of $\gamma_1$ with respect to agents' relative $x-$position; purple represents low densities and yellow represents high density. \textbf{(b)} Estimated distribution of $\gamma_1$ from a responsibility CVAE with Gaussian latent space. \textbf{(c)} Agent 1's ground truth $x$-acceleration vs. $x$-accel. from conditionally sampled $\gamma_1$.}
\label{fig:synthetic results}
\vspace{-5mm}
\end{figure*}

Since we train our posterior to be conditioned on both states and controls, we also assume our dataset contains each agent $i$'s control at the conditioning trajectory's end-time $T$, denoted $u^{(T)}_i \in \mathbb{R}^m$. Each agent's control contains values representing its acceleration and/or steering commands. We define the sequence $\mathbf{u}$:
\begin{align*}
    \mathbf{u} = [u^{(T)}_1,\, u^{(T)}_2, \dots, \, u^{(T)}_N]^\top
\end{align*}
Similarly, the set of agent responsibilities $\gammavec$ is defined:
\begin{align*}
    \gammavec = [\gamma_1^{(T)}, \, \gamma_2^{(T)}, \, \dots, \gamma_N^{(T)}]^\top
\end{align*}
So, $\mathbf{u} \in \mathbb{R}^{N \times m}$ and $\gammavec \in \mathbb{R}^{N}$, with $\gammavec$ having the same form as in our original CVAE formulation. It is important to note that we only care about our target variable (each agent $i$'s control $u_i$) at the time of inference $T$ because we are performing just single-step prediction of $\gammavec$ at time $T$. This is due to $\gammavec$ being dependent on $\mathbf{s}$ which is not explicitly modeled. As a result, predicted sequences of $\gammavec$ without access to future $\mathbf{s}$ would be difficult to interpret.
Having defined $\mathbf{s}_{\text{past}}$, $\mathbf{u}$ and $\gammavec$, we can see our dataset $\mathcal{D}$ referenced in Eq. \eqref{eq:final-elbo} is still a collection of state-control pairs $(\mathbf{s}_{\text{past}}, \, \mathbf{u})$, but as variable-length sequences.

\subsection{Responsibility Transformer}
The transformer architecture we use follows from \cite{YuanWengEtAl2021}; while Fig. \ref{fig:responsibility cvae diagram} shows just a high-level CVAE diagram, it being a transformer encoder/decoder does not meaningfully change the process. In the encoder block, we pass $\mathbf{s}_{\mathrm{past}}$ and $\mathbf{u}$ through time-embedding and self-attention layers; these are both passed through the posterior transformer layers $q_\varphi$, while only $\mathbf{s}_{\mathrm{past}}$ is used for $p_\psi$. We then have latents $\mathbf{z} = [z_1, z_2, \dots, z_N]$, which are passed through more transformer layers (along with an embedded $\mathbf{s}_{\mathrm{past}}$), producing the estimated $\gammavec$ for the current time. Finally, these $\gammavec$ and the dynamics-related states from the current time produce estimates of $\mathbf{u}$ via Problem \ref{prob:responsible multiagent CBF safety filter}. Since training is batched across samples, we also set a maximum number of agents $N_{\max}$ and a maximum time horizon $T_{\max}$, meaning all inputs and outputs of the CVAE have fixed sizes with padding. Further, like any other transformer model, we can do autoregressive trajectory predictions based on prior responsibility predictions; we discuss this further in Sec. \ref{subsubsec: real eval}.

\section{Experiments and Discussion}
The first objective of our experiments (Sec. \ref{subsec:synthetic dataset}) is to show that we can reconstruct the underlying relationship between agents' contexts and their responsibilities with our probabilistic formulation compared to a synthetic ground truth. The second objective (Sec. \ref{subsec:real dataset}) is to show that our approach learns meaningful responsibilities for complex, time-varying scenarios with many agents moving in and out of the scene. Although trajectory forecasting is not our focus, we evaluate responsibility-induced trajectories in our ablation study to determine which responsibility model is most useful.
\subsection{Synthetic Dataset}
\label{subsec:synthetic dataset}
We first trained an MLP-based CVAE (rather than using transformers) on synthetic data to verify our probabilistic responsibility formulation. Hence, our state data had a history of $1$. We evaluate the CVAE's ability to reconstruct $p(\gammavec \mid \mathbf{s})$ and the induced control distribution.

\subsubsection{Dataset generation}
In this setup, two agents, both with linear double-integrator dynamics, approach each other in a ``corridor problem'' scenario. Both agents have desired controls of constant acceleration and must obey a collision avoidance constraint. We generate a bimodal conditional Gaussian distribution $p(\gammavec \mid \mathbf{s})$, where $\mathbf{s}$ are the agents' relative positions/velocities. $p(\gammavec \mid \mathbf{s})$ was generated so that as the agents' $x$-distance increases, their responsibility allocation becomes slightly more skewed, with a $50\%$ chance of one allocation vs. its opposite (eg. $50\%$ of mean $\bar \gammavec = [0.8, 0.2]$ vs. $50\%$ of $\bar \gammavec = [0.2, 0.8]$). Our generated $p(\gammavec \mid \mathbf{s})$ is visualized for agent 1 in Fig. \ref{fig:synthetic true gamma x}. Then, we created a dataset of state-control pairs by solving Problem \ref{prob:responsible multiagent CBF safety filter} using generated $(\gammavec, \mathbf{s}, \mathbf{u}^{\mathrm{desired}})$ samples; a cross-section of the ground truth control distribution is shown in blue in Fig. \ref{fig:synthetic u}.
\subsubsection{Training}
We trained two CVAE models, one with a continuous latent space and another with a discrete latent space. For both models, the encoder and decoder are composed of MLPs of three hidden layers with widths of $16$ and $\tanh$ activation functions. For the discrete CVAE, we used a latent size of $2$ and $2$ categories. For the continuous CVAE, we also had a latent size of $2$. We trained the discrete model for $20$ epochs and the continuous model for $40$ epochs, each with ADAM \cite{KingmaBa2015}, a learning rate of $10^{-3}$,  $\beta$-annealing of the KL regularization term \cite{HigginsMattheyEtAl2017}. For all experiments, our code is implemented in $\mathrm{JAX}$ \cite{JAX2018}, and we use $\mathrm{Equinox}$ \cite{KidgerGarcia2021}, a neural network library. To solve Problem \ref{prob:responsible multiagent CBF safety filter}, we use $\mathrm{qpax}$ \cite{TracyManchester2024}, a $\mathrm{JAX}$-based differentiable quadratic program solver.

\subsubsection{Results}
We found both models reconstruct the underlying relationship between $\mathbf{x}$ and $\gammavec$; however, we observed that the discrete CVAE's performance was sensitive to changes in the latent size and the number of categories, complicating hyperparameter optimization. Thus, we focus on the Gaussian CVAE's performance here and in the next set of experiments. As seen in Fig. \ref{fig:synthetic continuous gamma x} the Gaussian CVAE is able to recover the bimodal nature of $\gammavec$, and its generated control distribution (in orange) is accurate to the ground truth (Fig. \ref{fig:synthetic u}), validating our CVAE approach.

\begin{figure*}[t]
  \centering
  \includegraphics[width=\linewidth]{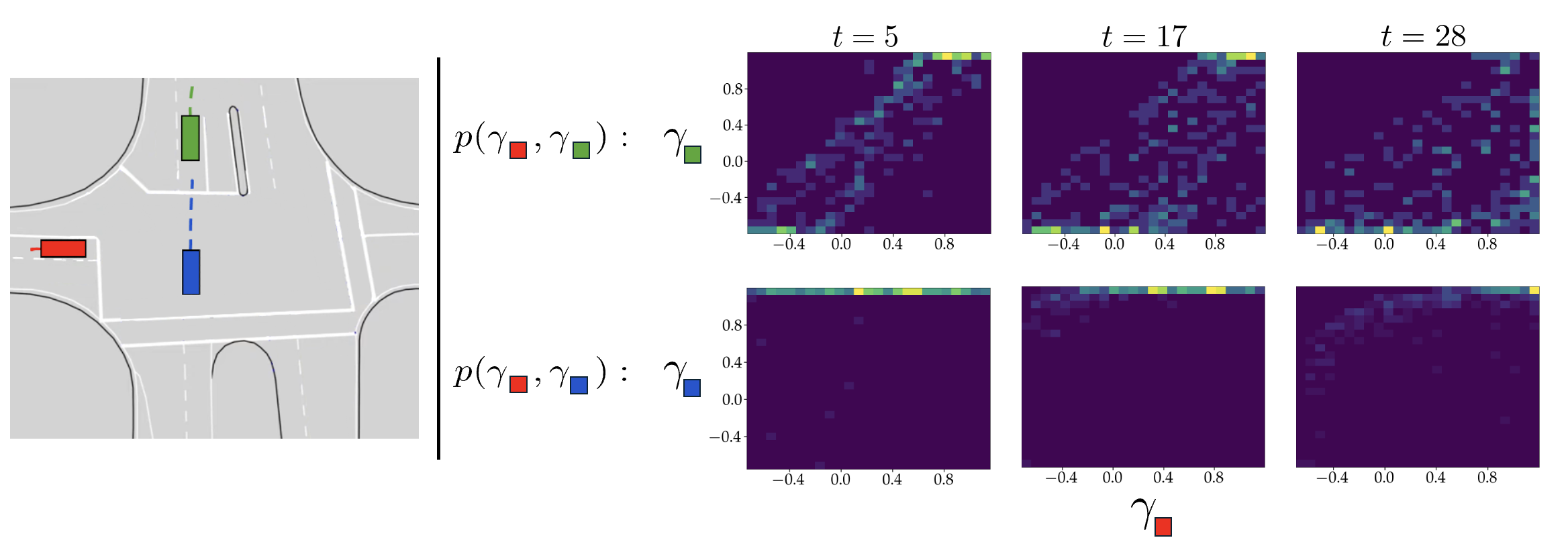}
  \vspace{-7mm}
  \caption{Responsibility allocations $\gammavec$ in a traffic intersection from the INTERACTION dataset. \textbf{Left:} The blue car arrives in the interaction first, and is leaving (dashed lines denote past trajectories), while the red car has arrived before green. \textbf{Right:} Estimated probability densities of $\gamma$ for red vs. green and red vs. blue over time; more yellow means higher density.}
\label{fig:gamma intersection}
\vspace{-4mm}
\end{figure*}

\subsection{INTERACTION Dataset}
\label{subsec:real dataset}
We use the INTERACTION dataset \cite{ZhanSunEtAl}, which contains human driving trajectories collected via drone (sampled at $10$Hz), to test the scalability of our probabilistic responsibility model. We use five minutes of trajectory data, yielding $30$k datapoints where we evaluate responsibility (with an $80/20$ train-test split), from one multi-way intersection scene. Different cars are constantly arriving, yielding and leaving, as shown in Fig. \ref{fig:gamma intersection} on the left.

\subsubsection{Desired control policy}
To retrieve the desired control, we use a linear feedback policy that uses an agent's current velocity and an assumed desired velocity (the maximum velocity in this scene, $~10\text{m/s}$, which is achieved when agents exit the intersection) to get the desired acceleration. We assume each agent follows linear double integrator dynamics.

\subsubsection{Model setup}
We constructed the transformer-based CVAE previously described in Sec.~\ref{sec:transformer model}, with a Gaussian latent space for $\mathbf{z}$, maximum history $T_{\max} = 30$ ($3$ seconds) and maximum number of agents $N_{\max} = 6$. Since there are many scenes where more than $N_{\max}$ agents are present, agents are selected by picking an ``ego'', and then the $N_{\max}-1$ closest agents (in L2 distance) to the ego. Our model uses $4$ transformer layers with $8$ attention heads each, an embedding size of $64$, hidden-layer widths of $256$, and a latent $\mathbf{z}$ size of $32$. Each model we tested was trained for $100$ epochs with batch sizes of $32$, a learning rate of $10^{-3}$, and $\beta$-annealing. With these parameters, a model has, responsibility safety filter included, an approximate inference time of $0.463_{\pm 8.72\text{e-}3}$ seconds (with a sample size of $32$).

\subsubsection{Evaluation}
\label{subsubsec: real eval}
Note that we cannot evaluate how well the model predicts $\gammavec$ directly since there is no ground truth data on $\gammavec$. Instead, we use predicted $\gammavec$ to produce new controls $\mathbf{u}$ via Problem \ref{prob:responsible multiagent CBF safety filter}, integrate those controls through our dynamics from the current state, and repeat this process until we have a trajectory of multi-agent states, which we evaluate using standard trajectory forecasting methods: average displacement error (ADE) and miss rate (average true/false rate of whether the maximum distance error in predicted trajectories exceeds a threshold; ours is $1$m). 

\subsubsection{Ablations}
When inferring $\gammavec$, prior work \cite{RemyFridovichKeilEtAl2025} constrained $\gammavec$ such that $\mathbf{1}^\top\gammavec = 1$ and $\gamma_i \in [0, 1]$. However, we observe that any $\gamma_i$ can in theory be negative, incentivizing an agent $u_i$ to pursue its desired control \textit{even less}. Thus, we ablate over output activation functions to pass $\gammavec$ through to see how different constraints on $\gammavec$ affect model performance.

\begin{table}[t]
\setlength{\tabcolsep}{10pt}
\renewcommand{\arraystretch}{1.3}
\centering
\caption{Predicted trajectory performance on INTERACTION data over $1$ second prediction horizon. Lower is better.}
\label{tab:benchmark_comparison}
\begin{tabular}{|l|c|c|}
\hline
\textbf{Method} & \textbf{ADE (m)} & \textbf{Miss Rate (\%)} \\
\hline
Direct $\mathbf{u}$ prediction        & $0.140_{\pm 9.10\text{e-}3}$ & $ 1.27\%_{\pm 2.85\text{e-}4}$\\
Following $\mathbf{u}_{\mathrm{desired}}$& $0.406_{\pm 2.26\text{e-}3}$ & $49.3 \%_{\pm 7.27\text{e-}3}$\\
\hline
\textbf{(1)} \quad $\mathrm{none}$            & $ 0.221_{\pm 9.44\text{e-}3}$ & $9.31\%_{\pm 2.83\text{e-}3}$ \\
\textbf{(2)} \quad $\mathrm{softmax}$        & $\mathbf{0.178_{\pm 9.91\text{e-}3}}$ & $10.1\%_{\pm  3.47\text{e-}3}$ \\
\textbf{(3)} \quad $\mathrm{clip}$ at $0$            & $\mathbf{ 0.185_{\pm 9.76\text{e-}3}}$ & $9.80\%_{\pm 2.83\text{e-}3}$ \\
\textbf{(4)} \quad $\mathrm{clip}$ at $-\beta$            & $0.216_{\pm 9.30\text{e-}3}$ & $\mathbf{8.74\%_{\pm 2.87\text{e-}3}}$ \\
\textbf{(5)} \quad $\tanh$                         & $0.209_{\pm 9.38\text{e-}3}$ & $\mathbf{8.83\%_{\pm 2.77\text{e-}3}}$ \\
\hline
\end{tabular}
\vspace{-5mm}
\end{table}

\subsubsection{Quantitative results}
In Table \ref{tab:benchmark_comparison} we show trajectory prediction results for responsibility CVAEs trained using different activation functions on $\gammavec$, compared to two baselines (top): trajectories from a CVAE that predicts $\mathbf{u}$ directly (using the same CVAE architecture, but with no responsibility safety filter layer), and trajectories from the $\mathbf{u}_{\mathrm{desired}}$ heuristic. We train $p(\gammavec \mid \mathbf{s})$ with: \textbf{(1)} no output activation, \textbf{(2)} $\mathrm{softmax}(\gammavec)$, \textbf{(3)} $\mathrm{clip}(\gamma_i, 0, \infty)$, \textbf{(4)} $\mathrm{clip}(\gamma_i, -\beta, \infty)$, and \textbf{(5)} $\tanh(\gamma_i)$.

\subsubsection{Qualitative results}
For our discussion, we focus on an useful case-study from the responsibility model with a $\tanh$ output layer, visualized in Fig. \ref{fig:gamma intersection}. This involves three cars at the intersection---the red car slowly moves forward while waiting for the blue car to finish crossing the intersection, and the green car is just arriving at the stop. 

\subsubsection{Takeaways}
Our quantitative and qualitative results lend three key takeaways:
\textbf{(i)} \textbf{Our approach allows for strong predictive performance across the board in trajectory prediction.} As Table \ref{tab:benchmark_comparison} shows, all five models vastly improve over the $\mathbf{u}_{\mathrm{desired}}$ heuristic---since the safety filter is more restrictive, it also makes sense that integrating direct $\mathbf{u}$ predictions has better performance, providing a lower bound.

\textbf{(ii)} \textbf{Some models are sensitive to high bias towards $0$ controls.} We found the INTERACTION data we use has agent controls (acceleration) that follow a narrow Gaussian centered at $\mathbf{u} = 0$, i.e., most agents are going at a constant velocity. As a result, the $\mathrm{softmax}$ and $\mathrm{clip(\gamma_i, 0, \infty)}$ models overfit to producing uniform responsibility allocations ($\gammavec = [0, 0, \dots, 0], \gammavec = [\frac{1}{N}, \frac{1}{N}, \dots, \frac{1}{N}]$ respectively) in all scenarios. So, these models have lower ADE while their miss-rate is worse---this is due to outlier cases where $\gamma_i < 0$ (less restrictive) induces more accurate controls leading to the $\mathrm{tanh}$ and $\mathrm{clip(\gamma_i, -\beta, \infty)}$ models finding trajectory modes that match the data. 

\textbf{(iii)} \textbf{We learn meaningful responsibilities.} In Fig. \ref{fig:gamma intersection} we see that for the red and green cars early on, there is bimodal $\gamma_{\colorbox{DarkGreen}{}}$, meaning the green car can either speed up or slow down, while $\gamma_{\colorbox{Red}{}}$ is roughly skewed-Gaussian centered at $0$; as green approaches the intersection, the distribution becomes more skewed towards negative $\gamma_{\colorbox{DarkGreen}{}}$ while $\gamma_{\colorbox{Red}{}}$ becomes more skewed towards $1$, lining up with our intuition that green would have more responsibility (smaller $\gamma_{\colorbox{DarkGreen}{}}$) by the end of this episode (as it arrived at the intersection after red).
For red and blue, we again see how red starts moderately skewed, and becoming more so as blue passes by, whereas blue retains a high $\gamma_{\colorbox{Blue}{}}$ throughout the whole episode---this also matches our intuition since blue was the first car at the intersection whereas red is waiting for it to pass. 

\section{Conclusions and future work}
In this work we presented a novel probabilistic formulation of multi-agent responsibility, taking an useful existing responsibility formulation and scaling it to complex scenes with multi-modal behavior and variable, arbitrary numbers of agents. Our CVAE approach demonstrated strong reconstruction performance on both synthetic and real-world data, and we showed how a learned responsibility CVAE can give predictions that match our own intuition in a real-world driving setting. However, we found the real-world dataset was heavily biased toward constant velocity scenarios, causing a lack of diversity in different responsibility outcomes. Further, we used a distance-based heuristic to filter for ``relevant'' agents, when a filter grounded in their dynamics might be better for finding zones of interaction.

Motivated by our findings, there are multiple avenues we wish to explore: \textbf{(i)} incorporating complex road geometry and visual context (e.g., road signs), more scenes, and more diverse bodies of agents (eg. pedestrians, bicycles), \textbf{(ii)} testing methods that better filter out irrelevant agents during responsibility prediction, like reachability-based methods, and \textbf{(iii)} investigating how learned responsibility models can lend insight into existing human or robot policies---perhaps through counterfactual scenarios, or through comparing prior and posterior responsibility predictions.

\bibliographystyle{IEEEtran}
\bibliography{main,ctrl_papers}

@Preamble{"\newcommand{\noopsort}[1]{} " #
"\newcommand{\printfirst}[2]{#1} " #
"\newcommand{\singleletter}[1]{#1} " #
"\newcommand{\switchargs}[2]{#2#1} "}

@String{jrn_IEEE_RAL                   = {{IEEE Robotics and Automation Letters}}}

@String{jrn_IEEE_TAC                   = {{IEEE Transactions on Automatic Control}}}

@String{jrn_IEEE_TIV                   = {{IEEE Transactions on Intelligent Vehicles}}}

@String{jrn_PNAS                       = {{Proceedings of the National Academy of Sciences}}}

@String{jrn_SAGE_IJRR                  = {{Int.\ Journal of Robotics Research}}}

@String{proc_ECCV                      = {{European Conf.\ on Computer Vision}}}

@String{proc_ICLR                      = {{Int.\ Conf.\ on Learning Representations}}}

@String{proc_IEEE_ACC                  = {{American Control Conference}}}

@String{proc_IEEE_CDC                  = {{Proc.\ IEEE Conf.\ on Decision and Control}}}

@String{proc_IEEE_CVPR                 = {{IEEE Conf.\ on Computer Vision and Pattern Recognition}}}

@String{proc_IEEE_ICCV                 = {{IEEE Int.\ Conf.\ on Computer Vision}}}

@String{proc_IEEE_ICRA                 = {{Proc.\ IEEE Conf.\ on Robotics and Automation}}}

@String{proc_IEEE_IROS                 = {{IEEE/RSJ Int.\ Conf.\ on Intelligent Robots \& Systems}}}

@String{proc_IEEE_ITSC                 = {{Proc.\ IEEE Int.\ Conf.\ on Intelligent Transportation Systems}}}

@String{proc_NIPS                      = {{Conf.\ on Neural Information Processing Systems}}}

@InProceedings{SchmerlingLeungEtAl2018,
  author    = {Schmerling, E. and Leung, K. and Vollprecht, W. and Pavone, M.},
  booktitle = proc_IEEE_ICRA,
  title     = {{Multimodal Probabilistic Model-Based Planning for Human-Robot Interaction}},
  year      = {2018},
  arxiv     = {1710.09483},
  category  = {interaction},
  img       = {SchmerlingLeungEtAl2018.png},
  selected  = {true},
}

@Article{IvanovicLeungEtAl2020,
  author   = {Ivanovic, B.* and Leung, K.* and Schmerling, E. and Pavone, M.},
  journal  = jrn_IEEE_RAL,
  number   = {2},
  pages    = {295--302},
  title    = {{Multimodal Deep Generative Models for Trajectory Prediction: A Conditional Variational Autoencoder Approach}},
  volume   = {6},
  arxiv    = {2008.03880},
  category = {interaction},
  img      = {IvanovicLeungEtAl2020.jpg},
  selected = {true},
  year     = {2021},
}

@InProceedings{CosnerChenEtAl2023,
  author    = {Cosner, R. and Chen, Y. and Leung, K. and Pavone, M.},
  booktitle = proc_IEEE_ICRA,
  title     = {{Learning Responsibility Allocations for Safe Human-Robot Interaction with Applications to Autonomous Driving}},
  year      = {2023},
  arxiv     = {2303.03504},
  category  = {structure},
  img       = {CosnerChenEtAl2023.png},
  selected  = {true},
}

@InProceedings{GeldenbottLeung2024,
  author    = {Geldenbott, J. and Leung, K.},
  booktitle = proc_IEEE_ICRA,
  title     = {{Legible and Proactive Robot Planning for Prosocial Human-Robot Interactions}},
  year      = {2024},
  arxiv     = {2404.03734},
  category  = {interaction},
  img       = {GeldenbottLeung2024.png},
  selected  = {true},
}

@InProceedings{HsuLeungEtAl2023,
  author    = {Hsu, K-C. and Leung, K. and Chen, Y. and Fisac, J. and Pavone, M.},
  booktitle = proc_IEEE_IROS,
  title     = {{Interpretable Trajectory Prediction for Autonomous Vehicles Via Counterfactual Responsibility}},
  year      = {2023},
  category  = {structure},
  img       = {HsuLeungEtAl2023.png},
  paper     = {https://ieeexplore.ieee.org/document/10341712},
}

@InProceedings{RemyFridovichKeilEtAl2025,
  author    = {Remy, I. and Fridovich-Keil, D. and Leung, K.},
  booktitle = proc_IEEE_ACC,
  title     = {{Learning responsibility allocations for multi-agent interactions: A differentiable optimization approach with control barrier functions}},
  year      = {2025},
  arxiv     = {2410.07409},
  category  = {structure},
  img       = {RemyFridovichKeilEtAl2024.png},
  selected  = {true},
}

@article{Doersch2016,
  author    = {Doersch, C.l},
  title     = {{Tutorial on variational autoencoders}},
  year      = {2016},
  journal      = {{Available at }\url{https://arxiv.org/abs/1606.05908}},
}

@InProceedings{JangGuEtAl2017,
  Title                    = {Categorial Reparameterization with Gumbel-Softmax},
  Author                   = {Jang, E. and Gu, S. and Poole, B.},
  Booktitle                = proc_ICLR,
  Year                     = {2017}
}

@Article{RudenkoPalmieriEtAl2020,
  author  = {Rudenko, A. and Palmieri, L. and Herman, M. and Kitani, K.~M. and Gavrila, D.~M. and Arras, K.~O.},
  journal = jrn_SAGE_IJRR,
  number  = {8},
  pages   = {895--935},
  title   = {{Human motion trajectory prediction: A survey}},
  volume  = {39},
  year    = {2020},
}

@Article{JAX2018,
  author  = {Bradbury, J. and Frostig, R. and Hawkins, P. and Johnson, M.~J. and Leary, C. and Maclaurin, D. and Necula, G. and Paszke, A. and Vander{P}las, J. and Wanderman-{M}ilne, S. and Zhang, Q.},
  journal = {Available at \url{http://github.com/google/jax}},
  title   = {{{JAX}: composable transformations of {P}ython+{N}um{P}y programs}},
  year    = {2018},
}

@Article{CaesarBankitiEtAl2019,
  author  = {Caesar, H. and Bankiti, V. and Lang, A. H. and Vora, S. and Liong, V. E. and Xu, Q. and Krishnan, A. and Pan, Y. and Baldan, G. and Beijbom, O.},
  journal = {{Available at } \url{https://arxiv.org/abs/1903.11027}},
  title   = {{{nuScenes}: A multimodal dataset for autonomous driving}},
  year    = {2019},
}

@Article{SchwartingPiersonEtAl2019,
  author  = {Schwarting, W. and Pierson, A. and Alonso-Mora, J. and Karaman, S. and Rus, D.},
  journal = jrn_PNAS,
  number  = {50},
  pages   = {24972--24978},
  title   = {{Social behavior for autonomous vehicles}},
  volume  = {116},
  year    = {2019},
}

@InProceedings{SunZhanEtAl2018,
  author    = {Sun, L. and Zhan, W. and Tomizuka, M. and Dragan, A.},
  booktitle = proc_IEEE_IROS,
  title     = {{Courteous Autonomous Cars}},
  year      = {2018},
}

@InProceedings{ToghiValienteEtAl2021,
  author    = {Toghi, B. and Valiente, R. and Sadigh, D. and Pedarsani, R. and Fallah, Y. P.},
  booktitle = proc_IEEE_IROS,
  title     = {{Cooperative Autonomous Vehicles that Sympathize with Human Drivers}},
  year      = {2021},
}

@Article{ToghiValienteEtAl2022,
  author  = {Toghi, B. and Valiente, R. and Sadigh, D. and Pedarsani, R. and Fallah, Y.},
  journal = jrn_IEEE_TIV,
  number  = {12},
  pages   = {24791--24804},
  title   = {{Social Coordination and Altruism in Autonomous Driving}},
  volume  = {23},
  year    = {2022},
}

@InProceedings{SunKretzschmarEtAl2020,
  author    = {Sun, P. and Kretzschmar, H. and Dotiwalla, X. and Chouard, A. and Patnaik, V. and Tsui, P. and Guo, J. and Zhou, Y. and Chai, Y. and Caine, B. and Vasudevan, V. and Han, W. and Ngiam, J. and Zhao, H. and Timofeev, A. and Ettinger, S. and Krivokon, M. and Gao, A. and Joshi, A. and Zhang, Y. and Shlens, J. and Chen, Z. and Anguelov, D.},
  booktitle = proc_IEEE_CVPR,
  title     = {{Scalability in Perception for Autonomous Driving: Waymo Open Dataset}},
  year      = {2020},
  month     = {June},
}

@InProceedings{AmesGrizzleEtAl2014,
  author    = {Ames, A.~D. and Grizzle, J.~W. and Tabuada, P.},
  booktitle = proc_IEEE_CDC,
  title     = {{Control barrier function based quadratic programs with application to adaptive cruise control}},
  year      = {2014},
}

@Article{AmesXuEtAl2017,
  author  = {Ames, A.~D. and Xu, X. and Grizzle, J.~W. and Tabuada, P.},
  journal = jrn_IEEE_TAC,
  number  = {8},
  pages   = {3861--3876},
  title   = {{Control Barrier Function Based Quadratic Programs for Safety Critical Systems}},
  volume  = {62},
  year    = {2017},
}

@InProceedings{KingmaBa2015,
  author    = {Kingma, D. and Ba, J.},
  booktitle = proc_ICLR,
  title     = {{Adam: A Method for Stochastic Optimization}},
  year      = {2015},
}

@Article{TracyManchester2024,
  author  = {Tracy, K. and Manchester, Z.},
  journal = {{Available at } \url{https://arxiv.org/abs/2406.11749}},
  title   = {{On the Differentiability of the Primal-Dual Interior-Point Method}},
  year    = {2024},
}

@InProceedings{HigginsMattheyEtAl2017,
  author    = {Higgins, I. and Matthey, L. and Pal, A. and Burgess, C. and Glorot, X. and Botvinick, M. and Mohamed, S. and Lerchner, A.},
  booktitle = proc_ICLR,
  title     = {{beta-{VAE}: Learning Basic Visual Concepts with a Constrained Variational Framework}},
  year      = {2017},
}

@InProceedings{KidgerGarcia2021,
  author    = {Kidger, P. and Garcia, C.},
  booktitle = proc_NIPS,
  title     = {{{E}quinox: neural networks in {JAX} callable {P}y{T}rees and filtered transformations}},
  year      = {2021},
}

@InProceedings{ChenSinavskiEtAl2024,
  author    = {Chen, L. and Sinavski, O. and H\"{u}ermann, J. and Karnsund, A. and Willmott, A.~J. and Birch, D.},
  booktitle = proc_IEEE_ICRA,
  title     = {{Driving with LLMs: Fusing Object-Level Vector Modality for Explainable Autonomous Driving}},
  year      = {2024},
}

@Article{MaoQianEtAl2023,
  author  = {Mao, J. and Qian, Y. and Ye, J. and Zhao, H. and Wang, Y.},
  journal = {{Available at } \url{https://arxiv.org/abs/2310.01415}},
  title   = {{GPT-Driver: Learning to Drive with GPT}},
  year    = {2023},
}

@InProceedings{ZhouHuangEtAl2024,
  author    = {Zhou, Y. and Huang, L. and Bu, Q. and Zeng, J. and Li, T. and Qiu, H. and Zhu, H. and Guo, M. and Qiao, Y. and Li, H.},
  booktitle = proc_ECCV,
  title     = {{Embodied Understanding of Driving Scenarios}},
  year      = {2024},
}

@InProceedings{RhinehartMcAllisterEtAl2019,
  author    = {Rhinehart, N. and McAllister, R. and Kitani, K. and Levine, S.},
  booktitle = proc_IEEE_ICCV,
  title     = {{PRECOG: PREdiction Conditioned On Goals in Visual Multi-Agent Settings}},
  year      = {2019},
}

@Article{TangMa2025,
  author  = {Tang, Y. and Ma, W},
  journal = {{Available at }\url{https://arxiv.org/abs/2503.04952}},
  title   = {{INTENT: Trajectory Prediction Framework with Intention-Guided Contrastive Clustering}},
  year    = {2025},
}

@Article{XuYangEtAl2025,
  author  = {Xu, Y. and Yang, R. and Zhang, Y. and Wang, Y.},
  journal = {{Available at }\url{https://arxiv.org/abs/2506.03408}},
  title   = {{Trajectory Prediction Meets Large Language Models: A Survey}},
  year    = {2025},
}

@Article{KingmaWelling2022,
  author  = {Kingma, D.~P. and Welling, M.},
  journal = {{Available at } \url{https://arxiv.org/abs/1312.6114}},
  title   = {{Auto-Encoding Variational Bayes}},
  year    = {2022},
}

@InProceedings{YuanWengEtAl2021,
  author    = {Yuan, Y. and Weng, X. and Ou, Y. and Kitani, K.},
  booktitle = proc_IEEE_ICCV,
  title     = {{AgentFormer: Agent-Aware Transformers for Socio-Temporal Multi-Agent Forecasting}},
  year      = {2021},
}

@InProceedings{LyuLuoEtAl,
  author    = {Lyu, Y. and Luo, W. and Dolan, J.M.},
  booktitle = proc_IEEE_ITSC,
  title     = {{Responsibility-associated Multi-agent Collision Avoidance with Social Preferences}},
  year      = {2022},
}

@Article{ZhanSunEtAl,
  author  = {Zhan, W. and Sun, L. and Wang, D. and Shi, H. and Clausse, A. and Naumann, M. and K\"ummerle, J. and K\"onigshof, H. and Stiller, C. and de La Fortelle, A. and Tomizuka, M.},
  title   = {{INTERACTION} {Dataset}: {An} {INTERnational}, {Adversarial} and {Cooperative} {moTION} {Dataset} in {Interactive} {Driving} {Scenarios} with {Semantic} {Maps}},
  journal = {Available at {https://arxiv.org/abs/1910.03088}},
  year    = {2019},
}

\end{document}